\documentstyle[12pt,epsfig]{article}

\begin{document}

\title{Quantum TAP equations}
\vskip 10pt
\author{Giulio Biroli$^{1,2}$ and Leticia F. Cugliandolo$^{2,3}$ 
\\ \\{\small
$^1$Center for Material Theory, Department of Physics and 
Astronomy}, \\
{\small Rutgers University, Piscataway, NJ 08854 USA} \\
{\small $^2$Laboratoire de Physique Th{\'e}orique de l'Ecole Normale 
Sup{\'e}rieure\thanks{Unit{\'e} Mixte de Recherche du Centre National de la 
Recherche Scientifique et de l'Ecole Normale Sup{\'e}rieure.}}
\\ {\small24 rue Lhomond, 75231 Paris cedex 05, France.} \\
{\small$^3$Laboratoire de Physique Th{\'e}orique et Hautes Energies, Jussieu, }\\
{\small4, Place Jussieu, 75252 Paris Cedex 05, France }}

\maketitle

\begin{abstract}
We derive Thouless-Anderson-Palmer (TAP) equations for
quantum disordered systems. We apply them to the study of 
the paramagnetic and glassy phases in 
the quantum version of the spherical $p$ spin-glass model. 
We generalize several useful quantities (complexity, threshold level, etc.)
and various ideas (configurational entropy crisis, etc),
that have been developed within the classical TAP approach,
to quantum systems. The analysis of the quantum TAP equations 
allows us to show that the phase diagram (temperature-quantum parameter)
of the $p$ spin-glass model should be generic. 
In particular, we argue that a crossover from a second order thermodynamic
transition close to the classical critical point to a 
first order thermodynamic transition close 
to the quantum critical point is to be expected in a large 
class of systems. 

\end{abstract}
\vspace{2cm}
\noindent
LPT-ENS/0033, LPTHE/0034.

\newpage
\section{Introduction}
\label{intro}

Glassy systems of extremely diverse types exist in nature. 
They all share 
several common features like a very slow, non-equilibrium dynamics.
The development of a full theoretical description of 
the glassy phase 
is one of the most important challenges in condensed matter physics. 
A variety of techniques, that range from scaling arguments to 
mean-field approaches have been, and are still used, with the 
aim of attempting a satisfactory description of the glassy properties.  

One of these techniques is due to 
Thouless, Anderson and Palmer (TAP) \cite{TAP}, who introduced an approach 
to classical disordered systems
based on the study of a free-energy landscape. The key object
is the Legendre transform of 
the free-energy $F(\beta) =-\ln Z / \beta $ with respect to 
a number of order parameters that are sufficient to 
describe the transition and the different phases in the system.
This function behaves as an effective potential whose minima 
represent  different possible phases. 
In a classical fully-connected Ising model only one order parameter is needed,
the global  magnetization $m=\sum_i \langle s_i\rangle/N$. The two possible 
minima of $F(\beta,m)$ correspond to the two possible 
states of positive and negative magnetization, $m = \pm m_o(T)$. 
Focusing on the Sherrington-Kirkpatrick (SK) mean-field model for 
spin-glasses, TAP showed that {\it all} the local magnetizations 
$m_i=\langle s_i\rangle$, $i=1,.\dots,N$, have to be included 
in order to derive the relevant free-energy landscape.
The extremization condition of the TAP free-energy
on the $m_i$s leads to the TAP equations. It was soon after realized
by Bray and Moore \cite{deDo,Brmo_TAP} that the number of solutions to the 
TAP equations for the SK model  
is exponential in  the number of spins in the system,
for temperatures below the spin-glass transition \cite{Ku}.
A very useful alternative derivation of the TAP equations 
was given by Plefka \cite{Pl} who showed that these equations can 
also be obtained from a power expansion of the Gibbs potential up to 
second order in the exchange couplings. The  advantage of this derivation 
is twofold:  it allows to show convergence of the power expansion 
for all temperatures and it is easily applicable to other  
mean-field glassy models. Moreover, Georges and Yedidia \cite{Geye}
showed that the high temperature series, at fixed order parameter,
of the free-energy can be used to derive TAP-like equations, 
and its corrections, 
for models in finite dimensions or, equivalently, with finite 
range interactions.
The connection between the TAP approach and the more standard analysis of the 
partition function of a disordered model has been  exhibited by 
De Dominicis and Young \cite{deDoyo} who showed that, for the SK model, one 
recovers the equilibrium results of the 
replica or the cavity method \cite{Beyond} via weighted Boltzmann averages over solutions
of the TAP equations. 
More recently, the TAP approach has been applied to other classical disordered models. 
In particular, two models that we shall discuss in the following, 
the spherical and Ising $p$ spin-glass models 
\cite{Kiwo-TAP,Rieger,Kupavi,Crso2,Cagipa1,Cagagi} and the 
Ghatak-Sherrington (GS)
model \cite{Gash,Mosh}  have been analyzed with this method \cite{Yo,Reop,deAr3}.

Glassy systems, and in particular disordered ones, are characterized by having 
a very slow dynamics with non equilibrium effects at low 
temperatures \cite{Bocukume,glassy_exps}. Mean-field models, 
like the the  spherical $p$ spin-glass model \cite{Cuku} 
or the SK spin-glass \cite{Cuku2}, 
capture this phenomenology. The dynamic solution for the evolution 
starting from random initial conditions, that represent 
a quench from high temperatures analytically, is intimately 
connected to the structure
and organization of TAP solutions. One of the most striking results of the 
dynamic analysis of $p$ spin-glass-like models 
is that the energy density (and other one time-quantities)
converges asymptotically to the energy density of high lying solutions 
of the TAP equations. This level has been called {\it threshold}. 
The  energy-density in equilibrium is different.
This and other related results suggest  that an interpretation of the 
dynamics  in terms of a motion in a TAP free-energy 
landscape can be given \cite{Cuku}.
The generalization of the TAP approach to dynamics that has been developed in 
\cite{Bi} allows one to make this statement precise: the evolution 
is determined by a gradient-descent in the TAP free-energy landscape with the 
most important addition of non-Markovian terms. 

Usually, glasses can be analyzed with a fully classical approach since 
their transition temperatures are rather high. Nevertheless, in 
many cases of great interest the critical temperature can be lowered 
by tuning another external parameter and quantum fluctuations become
very important. This is the case for the insulating magnetic compound
LiHo$_x$Y$_{1-x}$F$_4$, that is an experimental realization of a quantum spin-glass,
and presently receives much attention \cite{Ro}. Other examples where glassy 
properties in the presence of quantum fluctuations have been observed 
are mixed hydrogen bonded ferro-antiferro electric crystals \cite{exp_q2}, 
interacting electron systems \cite{Zvi}, cuprates
like La$_{2-x}$Sr$_x$CuO$_4$ \cite{cuprates}, 
amorphous insulators \cite{Osheroff}, etc.  

The quantum fluctuations in LiHo$_x$Y$_{1-x}$F$_4$ can be 
controlled by tuning the strength of an external field that 
is transverse to the 
preferred direction of the randomly located magnetic impurities. 
After a series of experiments presented in \cite{Ro} the authors' 
conclusions are: 
 (1) The samples undergo a paramagnetic to spin-glass transition in the 
$(T,\Gamma)$ plane, where $\Gamma\propto H_t^2$  and $H_t$ is 
the strength of the 
transverse field. 
(2) The transition is of second order (in the thermodynamic sense) close to the classical
critical point $(T=T_c,\Gamma=0)$ but crosses over to first order close to the 
quantum critical point   $(T=0,\Gamma=\Gamma_c)$. 
(3) The system undergoes out of equilibrium 
dynamics in the glassy phase as demonstrated by the fact that 
the dynamics strongly depends on the 
preparation of the sample for all subsequent times explored experimentally. 

The theoretical study of quantum spin-glasses  started with Bray and Moore's
analysis of the equilibrium properties of the 
fully connected Heisenberg model \cite{Brmo}. In this article, Bray and Moore
introduced a path-integral representation in imaginary time of the 
partition function that they analyzed with the replica trick. 
Many articles on the equilibrium of this, and related, 
mean-field models have been 
published since 
\cite{quantum_mf,Isya,rotors,Cesare,Cesare1,Niri,Culo,Cugrsa,Chamon}. 
The static 
properties of low dimensional 
models have been studied and it has been shown that, in finite dimensions,
Griffiths-McCoy 
singularities are very important close to the quantum critical 
point \cite{quantum_finite}. In all these models, the transition from 
the paramagnetic to the spin-glass phase has been reported to be of second
order throughout. 

In most classical disordered models studied so far the transition from
the disordered to the ordered phase is of second order in the 
thermodynamic sense. In the exact solution of the SK model, 
the spin-glass order parameter $q(x)$ is continuous at the transition which
is of second order in the thermodynamic sense \cite{Beyond}. 
In other classical glassy models like the Potts glass \cite{Kiwo-TAP} 
or the spherical \cite{Crso1} and Ising \cite{De,Kith} $p$ spin-glasses, 
the order parameter jumps at the transition which,
however, is still of second-order in the thermodynamic sense
since there is neither a jump in the susceptibility nor a 
latent heat. A classical model that exhibits a first-order transition
is the anisotropic $p$ spin-glass, $p\geq 3$, in 
which the spins take integer values between $-S$ and $S$ and 
there is an extra term in the Hamiltonian $-D\sum_i s_i^2$, proportional to a 
coupling constant $D$, that controls the crystalline tendency. 
In this case, a crossover from a second-order to a 
first-order thermodynamic transition in the plane $(T/J,D/J)$ has 
been exhibited in the exact solution \cite{Mo}. The classical Ghatak-Sherrington (GS) model \cite{Gash} is 
another candidate to 
exhibit a second to first order crossover in the thermodynamic transition. 
It is the anisotropic extension of the SK model, or the $p=2$ 
limit of the previous model.
In this case, a crossover from a second order  to a 
first order transition in the plane $(T/J,D/J)$ has been exhibited in 
an {\it approximate} solution (one step replica 
symmetry breaking) \cite{Gash,Mosh}. 
The exact solution has not been derived
yet and it is then not well established if it has 
a true first-order thermodynamic transition. 

In quantum problems, first order transitions have been 
reported in three models.
The first one is the so-called ``fermionic Ising spin-glass''
analyzed by Oppermann and collaborators \cite{Opper}. This model, however, 
is thermodynamically equivalent to the classical GS model discussed 
above \cite{Opper1}. 
The other two models are very similar indeed and 
they are different ways of extending the classical spherical $p$ spin-glass 
model \cite{Crso1} to include quantum fluctuations. 
In one case, the continuous spins are generalized to $M$ component vectors and 
a global spherical constraint as well  as commutation relations are imposed
\cite{Niri}.
The other one uses the fact that the spherical $p$ spin-glass model 
can be interpreted as a particle moving in an infinite 
dimensional hyper-sphere with a random potential. Quantization is then done 
by imposing commutation relations between coordinates and momenta \cite{Culo,Cugrsa}.
The latter can also be interpreted as an extension of the a quantum rotor model \cite{rotors}
that includes $p$ interactions. 
The relation between the critical properties of the quantum versions  of 
$p$  spin-glass models and the experiments in \cite{Ro} has been 
put forward in \cite{Cugrsa}. In addition,  
the connection between the static calculation 
supplemented by the marginality condition and the analysis of the 
out of equilibrium dynamics in contact with an environment 
developed in \cite{Culo} was also discussed in \cite{Cugrsa}.  
However, the reason why the transition changes from second to 
first order close to the 
quantum critical point was not clear from this analysis. It is one of the aims of this 
article to clarify this point, and study to what extent one can claim
it to be general, with the use of the TAP approach.

Quantum TAP equations for the SK model in a transverse field have been 
presented by Ishii and Yamamoto \cite{Isya} and Cesare {\it et al} 
\cite{Cesare1}. The former use a perturbative expansion  
of the free-energy 
in the strength of the transverse field,
and then follow closely TAP's techniques;
the latter implement a 
cavity method. The TAP equations derived 
by Rehker and Oppermann \cite{Reop}
for the fermionic spin-glass model coincide with the ones 
presented by Yokota \cite{Yo} for the classical GS model  
since these two models are thermodynamically equivalent \cite{Opper1}.

Hence, our aim is twofold. On the one hand we want to present a quantum extension of the 
TAP approach to the statistical properties of disordered systems. Thus, after
a short revision of the classical TAP approach in Section \ref{introtap},  
we discuss in Section \ref{sec1:formalism} the derivation of the quantum TAP free-energy and TAP equations using a 
general approach that extends the ones developed by Plefka \cite{Pl} and Georges and 
Yedidia \cite{Geye}. 
The advantage with respect to previous derivations of quantum TAP equations 
is that this method can be applied to any quantum disordered model
and it allows to obtain the TAP equations as well as the TAP free energy. 
In Section~\ref{p-spin} we present, as an example, the TAP free-energy and 
TAP-equations for the quantum extension of the $p$ spin spherical 
spin-glass model studied in \cite{Cugrsa,Culo}. We show that the TAP 
equations can be easily related to the equations for the order parameter
in the Matsubara replica approach and also to some of the equations 
appearing in the real-time dynamic approach. The TAP 
analysis of this model furnishes
 a benchmark to study the generalization to the quantum case 
of the methods and interpretations developed for classical systems.   
  Section~\ref{transition} is devoted to the second aim of this article. 
Via the TAP approach we show that the same type of phase diagram 
naturally emerges
for all systems having a discontinuous phase transition in their
classical limit (these are models solved by 
a one-step replica symmetry breaking Ansatz within the 
 replica analysis). In particular we relate the first-order 
transition close to the quantum critical point 
to the structure of metastable states. 
Finally, we present our conclusions in Section~\ref{conclu}.

\section{The classical TAP equations: a short revision}
\label{introtap}

In this section we present a short revision of the classical TAP approach
to mean-field disordered spin models. 
The classical TAP free-energy \cite{TAP} is the Legendre transform
of the free-energy with respect to local magnetic fields,
\begin{equation}\label{tapcl1}
-\beta F (\beta,m_{i})
=
{\mbox{Tr}}\exp \left(-\beta H -\sum_{i}h_{i}(s_{i}-m_{i})  \right)
\; ,
\end{equation}
where $h_{i}$ are Lagrange parameters enforcing the condition $\langle s_{i}\rangle=m_{i}$. $-\beta F(\beta,m_i)$ is an effective potential 
that depends on the local magnetizations. 
The Lagrange conditions $-\partial \beta F /\partial m_{i}=h_{i}$, called TAP
equations, fix the local magnetizations as functions of the local
magnetic fields 
\footnote{Note that within the TAP approach one does not average over disorder
from the beginning as in the replica method. Consequently the TAP free energy
 depends on the particular realization of disorder.}.
The solutions $\{m_{i}^{\alpha } \}$ of the TAP equations
 are stationary points of $F (\beta,m_{i})$. If they are also stable 
(all the corresponding eigenvalues of the free-energy Hessian are positive),
they are identified \cite{Beyond} with pure states, also called TAP 
states. This interpretation was put forward by 
De Dominicis and Young \cite{deDoyo} who
 showed that the partition function in the classical SK model 
can  be written as a weighted sum over the stable solutions of the TAP equations:
\begin{equation}
Z=\sum_{\alpha} \exp(-\beta F(\beta,{\bf m}^\alpha))
\; ,
\label{eq:weighted}
\end{equation}
where the index $\alpha$ labels different TAP states, ${\bf m}^\alpha$ is an 
$N$-vector encoding the local magnetization in the solution $\alpha$, 
$F$ is the extensive TAP free-energy of such solution and the sum runs 
over all TAP solutions.   
Consequently, the static average of any observable can be computed from 
Eq.~(\ref{eq:weighted}).
At low temperatures the TAP free-energy has a large number of minima. If one groups 
different TAP states with the same free-energy in sets ${\cal C}$ then
the partition function can be written as
\begin{equation}
Z=\sum_{{\cal C}} 
{\cal N}(f,\beta ) \exp(-\beta N f)
\label{eq:weighted1}
\end{equation}
where the factor ${\cal N}(\beta,f )$
is the number of solutions with TAP free-energy density
\newline 
$F(\beta,{\bf m}^\alpha)/N=f$.
One can now replace the sum by an integral
and  exponentiate the factor ${\cal N}(\beta,f )$; this yields 
\begin{equation}
\lim_{N\rightarrow \infty }\frac{1}{N}\ln Z=\lim_{N\rightarrow \infty }\frac{1}{N}\ln\int df \exp(-N (\beta f-\sigma(\beta,f  )))
\label{eq:weighted3}
\end{equation}
where we have taken the continuous limit and introduced the {\it complexity} 
\begin{equation}\label{complexity0}
\sigma(\beta,f) \equiv \lim_{N\to\infty} \frac{1}{N} 
\ln\left( {\cal N}(\beta,f) \right) 
\; .
\end{equation}
The configurations that dominate the sum are those having a
free-energy density such that it minimizes $\beta f - \sigma(\beta,f)$. 
The identity between the partition function and the weighted sum over TAP 
solutions has been demonstrated for many others models
 \cite{Crso2,Kiwo-TAP,Mon} and it is generally believed to hold  for any 
 mean-field disordered system. 

In the following we focus on ``discontinuous glassy systems'' 
\cite{Bocukume} that are characterized by having a discontinuous 
transition (the Edwards-Anderson order parameter, $q_{\sc ea}$, 
jumps) that is still of  
second order thermodynamically.
Within the replica analysis of the partition function 
these models are characterized by a one-step replica symmetry breaking
solution below a static transition temperature $T_s$ and a replica 
symmetric (RS) solution, that corresponds to the paramagnetic phase,
at $T> T_s$. However, for intermediate
temperatures $T_s < T < T_d$ there are an exponential in $N$
number of non-trivial TAP solutions that combine themselves in
such a way that the sum (\ref{eq:weighted3}) is identical to the 
RS result.

The relationship between metastable states and replicas has
been put forward in \cite{Kith,Mon,PaFr}. Indeed, consider
$x$ different identical systems (``clones'') coupled by an attractive,
infinitesimal (but extensive) interaction. When there exist
many pure states all the clones fall into the same state
and the free energy for the system of $x$ clones reads:
\begin{equation}\label{reltaprep0}
\lim_{N\rightarrow \infty }\frac{-1}{\beta N}\ln Z_{x}=\lim_{N\rightarrow \infty }\frac{-1}{\beta N}\ln \int df \exp(-N (\beta xf-\sigma(\beta,f)))
\; .
\end{equation}
On the other hand the computation of the left hand-side
 of (\ref{reltaprep0}) can be performed
within the replica formalism:
\begin{equation}\label{reltaprep1}
\lim_{N\rightarrow \infty }\frac{-1}{\beta N}\ln Z_{x}=\lim_{N\rightarrow \infty }\frac{-1}{\beta N}\overline{\ln Z_{x}}=\lim_{N\rightarrow \infty,n\rightarrow 0 }\frac{-1}{\beta N n}\ln \overline{Z_{x}^{n}}
\end{equation} 
where the overline represents the average over disorder. 
Since the attractive coupling between the $x$ clones is infinitesimal,
the computation of the right-hand side of (\ref{reltaprep1}) reduces
simply to the calculation of $\lim_{n'\rightarrow 0}(x/n')
\ln \overline{Z^{n'}}$,
where the replica symmetry between the $n$ groups of $x$-replicas ($n'=nx$)
is {\it explicitly} broken. When the system is in the replica symmetric phase ($T_{s}<T$),
this reduces to study one-step solutions non-optimized with respect to $x$:  
\begin{equation}\label{reltaprep}
-\lim_{N\rightarrow \infty}\frac{1}{\beta N }\ln \int df \exp(-N (\beta xf-\sigma(\beta,f)))
={x}{\mbox{Extr}}_{q_{\sc ea}}f_{\sc rep}(q_{\sc ea};\beta,x)
\end{equation}
where $f_{\sc rep}$ is the free-energy computed by using replicas, 
 $q_{\sc ea}$ is the Edwards-Anderson parameter and 
$x$ is the breakpoint (or the size of the blocks in the replica matrix). 
For simplicity we consider that the inter-state overlap 
$q_{0}$ equals zero. The definitions of these parameters are standard in the replica 
approach \cite{Beyond} and they will appear in the analysis of the 
quantum $p$ spin-glass model in Section~3. Since the integral on the 
left-hand-side of 
(\ref{reltaprep}) is dominated by a saddle point contribution, one finds
 that, for a given temperature, fixing the value 
 of $x$ is equivalent to summing over states with a 
given energy density $f$. The 
relationship between $f$ and $x$ reads
\begin{equation}\label{rela:xf}
\beta x=\frac{\partial\sigma(\beta,f) }{\partial f } 
\; .
\end{equation}  
Note that within
 this framework one does not have to optimize with respect to $x$.
Instead, $x$ is a free parameter and, by changing the value of $x$, 
one can consider different groups of metastable states.

The analysis of the TAP equations reveals three temperature regimes
 for discontinuous glassy systems: 
\begin{itemize}

\item 
{\it High temperatures} $T_d< T$. The system is in the paramagnetic phase, 
the paramagnetic 
TAP solution $m_i=0$, for all $i$, dominates the sum and $f_{pm} = -\ln Z/(\beta N)$.
$T_d$ is the dynamic critical temperature. Above $T_d$ the dynamics starting from a 
random initial condition converges asymptotically to the paramagnetic solution.

\item 
{\it Intermediate temperatures} $T_s <T< T_d$. The replica analysis of the partition function
indicates that the system is still in the paramagnetic phase. 
 However, the study of the TAP 
equations and the dynamics show that at $T_d$ the paramagnetic solution is 
fractured in an exponentially large in $N$ number of minima of the TAP
free-energy \cite{Kupavi,Kiwo-TAP, Alain,PaFr}.
 Indeed one can recover these results also by the replica method.
 A careful replica analysis shows that there exist one-step 
solutions in these temperature regime other than the paramagnetic one. These
 solutions are in one to
 one correspondence with groups of states with a given free-energy density
 (through the relationship (\ref{rela:xf})).
For instance, one can follow the evolution of the 
threshold states (the states with highest free-energy)
by tuning the parameter $x$. For these states, $x=1$ when $T=T_{d}$ and 
decreases at lower temperatures.
Moreover, the dominant contribution to 
Eq.~(\ref{eq:weighted3}) is given by the states characterized by $x=1$, 
{\it i.e.} those with free-energy density such that 
\begin{equation}
\beta=\frac{\partial\sigma(\beta,f)}{\partial f}
\; .
\end{equation}
These are the threshold states at $T_d$ and other groups when $T<T_d$.
Hence, between $T_s<T<T_d$ 
saddle-point solutions (corresponding to $x=1$) that are not absolute 
minima of $F$ dominate the integral since their number scales exponentially with $N$. The final result for the free-energy density 
in this temperature range coincides with the one of the prolongation 
of the paramagnetic solution (that actually does not exist!). 
A naive replica computation fails to signal the difference between a true 
paramagnetic solution and the ensemble of non trivial TAP 
solutions with $m_i\neq 0$. The dynamic approach detects 
the change in free-energy landscape at $T_d$ since the system 
cannot reach equilibrium for any temperature below $T_d$ \cite{Cuku}. 

\item 
{\it Low temperatures} $T<T_s$. At the static transition temperature the 
complexity of the 
TAP solutions, which dominates the sum (\ref{eq:weighted3}), vanishes. The 
static transition appears as an {\it entropy crisis}
since the part of the total entropy that is related to the large number 
of states disappears. For $T<T_{s}$ the TAP states which dominate
the integral in Eq.~(\ref{eq:weighted3}) 
correspond to the equilibrium glassy phase. Dynamically, $T_s$ does not play any role.
The out of equilibrium dynamics is dominated by the threshold states,
 which are the highest ones in free-energy and that are characterized 
 by flat directions in the free-energy landscape. 
\end{itemize}

Note that via the TAP approach one can obtain a reasonable justification of 
 the marginality condition \cite{replicon} often used to obtain information
 about the out of equilibrium dynamics starting from a pure equilibrium 
computation~\cite{Kith-replicon}.
Indeed the value of $x$ fixed by 
the marginality condition corresponds to the TAP states which are 
marginally stable (the threshold states):
 the flatness of the free-energy landscape around these
 states is responsible for aging \cite{Cuku}.

\section{The quantum TAP equations}
\label{sec1:formalism}

In this section we present a simple procedure
to derive TAP equations for generic completely connected 
quantum systems. We also expose the physical meaning
of quantum TAP equations by the cavity method \cite{Beyond}.

We are aware of two publications where TAP equations for quantum systems 
have been already presented \cite{Isya,Cesare1}.
With respect to these
works our derivation is more systematic, simple and  it
 allows one to obtain the TAP equations as well as the TAP free-energy for 
 any completely connected quantum disordered system.

\subsection{Formalism, notations and models}\label{introtapeq}

The formalism that we use to derive TAP equations for generic quantum problems 
is very similar to the one described in \cite{Pl,Geye} and, it 
follows even more closely, the one used in \cite{Bi} to obtain the 
dynamical TAP equations for classical disordered models.\\
We focus on systems characterized by the potential energy:
\begin{equation}\label{potential}
H_{p}=-\sum_{i_{1}<\dots <i_{p}}\sum _{\alpha }J_{i_{1},\dots ,i_{p}}s_{i_{1}}^{\alpha }
\cdots s_{i_{p}}^{\alpha } \qquad i=1,\dots ,N \quad \alpha =1,\dots ,m
\end{equation} 
where ${\mathbf{s}}_{i}$ may represent an SU(2) spin ($m=3$), 
a rotor ($m>1$), a spherical spin ($m=1$) or a space-coordinate ($m=1$)
 and $J_{i_{1},\dots ,i_{p}}$, the couplings between the different 
${\mathbf{s}}_{i}$, are independent random variables with zero mean and 
variance
\begin{equation}\label{coupling}
\overline{\left(J_{i_1...i_p}\right)^2} = {{\tilde{J}}^2 p!\over 2N^{p-1}}.
\end{equation}
As a consequence the following derivation applies to (completely connected) 
Heisenberg models,
quantum rotor models and quantum continuous systems\footnote{
The results obtained in this section can be straightforwardly
generalized to more complicate potentials containing different monomials 
or characterized by a more complicate tensorial coupling between 
$s_{i_{1}}^{\alpha _{1}}, \cdots , s_{i_{p}}^{\alpha _{p}}$.}.
Without loss of generality and to simplify the notation 
we shall suppress the index $\alpha $ in the rest of this section.

For classical spin-glasses TAP showed that all the local magnetizations 
$m_i$, $i=1,.\dots,N$, are needed to derive the relevant  free-energy
density to describe the metastable properties \cite{TAP}.
If one is interested in the dynamics of classical disordered 
mean-field systems, one has to Legendre transform  not only 
with respect to all 
time-dependent local magnetizations $m_i(t)$, but also 
with respect to 
the autocorrelation 
$C(t,t')=$
$1/N \sum_i \langle s_i(t) s_i(t')\rangle$
and the linear response   
$R(t,t')=1/N \sum_i \delta \langle s_i(t) \rangle/\delta h_i(t')|_{h=0}$ 
\cite{Bi}.

In order to describe the metastable properties of a  quantum disordered
model we shall show that it is necessary to Legendre transform 
with respect to the local average coordinates, $m_i(\tau)$, 
and the autocorrelation  function in imaginary time,
$C(\tau,\tau')$. 
The quantum TAP free-energy reads
\begin{eqnarray}
& & \left.-\beta F(\beta,m_i(\tau),C(\tau,\tau'),\alpha)
\right|_{\alpha =1} 
= 
\nonumber\\
& & 
\;\;\;\;\;\;\;\;\;\;
\ln 
\int {\cal D}{\bf s}(\tau) \,
\exp\left[ \;\; -\frac{1}{\hbar} \int_0^{\beta\hbar} d\tau \left(
H_{k} + \alpha H_{p}({\bf s}) \right)
\right.
\nonumber\\
& & 
\;\;\;\;\;\;\;\;\;\;
\left.
+
\frac{1}{2 \hbar^2} \int_0^{\beta\hbar} d\tau \int_0^{\beta\hbar} d\tau' \; \sum_{i}\Lambda (\tau,\tau') 
\left( C(\tau,\tau') - s_i(\tau) s_i(\tau') \right)
\right.
\nonumber\\
& & 
\;\;\;\;\;\;\;\;\;\;
+ 
\left.
\left.\frac{1}{\hbar} \int_0^{\beta\hbar} d\tau 
\sum_i h_i(\tau) ( m_i(\tau) - s_i(\tau)) 
\;\; \right]\right|_{\alpha =1} 
\; . 
\label{eq:Gamma}
\end{eqnarray}
where $H_{k}$ is the kinetic energy, ${\cal D}{\bf s}(\tau)$ indicates the
functional measure on the configuration space and $\alpha $
is a parameter whose role will be clarified in the following. For instance,
if ${\mathbf{s}}_{i}$ are SU(2) spins  $H_{k}$ is the Berry phase and 
the functional measure is restricted 
to periodic functions ${\mathbf{s}}_{i}(\tau )$ 
(with period $\beta $) satisfying the constraint 
${\mathbf{s}}^{2}_{i}(\tau )=1$.    
The sources $h_i(\tau)$ and $\Lambda (\tau,\tau')$ have the
 role of Lagrange multipliers fixing the average value of the coordinates and 
the correlation:
\begin{eqnarray}
m_i(\tau) &=& \langle s_i(\tau)\rangle 
\; ,
\\
C(\tau,\tau') &=& \frac{1}{N}\sum_{i}\langle s_i(\tau) s_i(\tau')\rangle
\; .
\end{eqnarray}
Once the TAP free-energy $F $ is known, one can derive the TAP equations
as Legendre relations,
\begin{equation}\label{legendre}
-\frac{\delta\beta F}{\delta m_{i}(\tau)} = h_{i}(\tau )\; ,
\qquad \qquad 
-\frac{2}{N} \frac{\delta\beta F}{\delta C(\tau,\tau')} = \Lambda(\tau,\tau')
\; .
\end{equation}
Until now we have not not used the scaling (\ref{coupling}) and all these definitions
can be equally applied to finite dimensional systems. The great simplification 
 due to the mean field character of the interactions in (\ref{coupling}) 
is unveiled if one performs a perturbative expansion of Eq.~(\ref{eq:Gamma})
in $\alpha $  and
writes
$-\beta F(\beta,m_i(\tau),C(\tau,\tau'),\alpha)$
as a power series in $\alpha$:
\begin{equation}
-\beta F(\beta,m_i(\tau),C(\tau,\tau'),\alpha) 
=
 \sum_{n=0}^\infty \frac{1}{n!} \left. 
\frac{\partial^n(-\beta F(\beta,m_i(\tau),C(\tau,\tau'),\alpha))}{\partial \alpha^n} \right|_{\alpha=0}
\alpha^n
\; .
\label{series}
\end{equation}
In fully-connected models, if one chooses the correct 
order parameters (which are $m_i(\tau)$, $C(\tau,\tau')$ 
in the quantum case), the perturbative expansion (\ref{series}) around the pure
 kinetic theory is actually a simple sum over three terms.
Higher order terms in the series vanish in the thermodynamic limit
due to the scaling of $J_{i_{1},\dots ,i_{p}}$ with respect to $N$. 
In more general cases, in finite dimensions, this will not be the case and 
(\ref{series}) becomes a $1/d$ expansion around mean field theory, where
$d$ is the spatial dimension \cite{Geye}.

Let us consider in more detail the terms arising from the expansion 
(\ref{series}). The zeroth-order one is simply $-\beta F(\beta,m_i(\tau),C(\tau,\tau'),0)$, i.e. the free-energy of $N$ free spins constrained to have 
local magnetizations $m_{i}(\tau )$ and a global correlation function
$C(\tau ,\tau ')$. This term depends only on the nature of the degrees 
of freedom, whether they are $SU(2)$ spins, rotors or space-coordinates.
In particular it can be analytically computed only if $s_{i}$ are 
spherical spins or space-coordinates. In the other cases one has to 
resort to approximations or numerical computations.

The first-order term is the naive mean field free-energy:
\begin{equation}\label{naivemfe}
\left. \frac{\partial (-\beta F)}{\partial \alpha} \right|_{\alpha=0}
=\frac{1}{\hbar}\int_{0}^{\beta \hbar }\sum_{i_{1}<\dots <i_{p}}J_{i_{1},\dots ,i_{p}}\left< 
s_{i_{1}}\cdots s_{i_{p}}\right>_{\alpha =0}=
\frac{1}{\hbar}\int_{0}^{\beta \hbar }\sum_{i_{1}<\dots <i_{p}}J_{i_{1},\dots ,i_{p}}
m_{i_{1}}\cdots m_{i_{p}}
\; .
\end{equation}
Note that the decoupling of the spins for $\alpha =0$ is essential to 
obtain the last identity. The second-order term depends on
the correlation function and the  overlap function 
$Q(\tau ,\tau ')=\sum_{i}m_{i}(\tau )m_{i}(\tau ')/N$ only and 
equals
\begin{equation}\label{secondterm}
\frac{N\tilde{J}^{2}}{4\hbar ^{2}} \int_0^{\beta \hbar } d\tau \int_0^{\beta \hbar } 
d\tau' \left( 
C^p(\tau,\tau') - Q^p(\tau,\tau') - p (C(\tau,\tau') - Q (\tau,\tau') ) 
Q^{p-1}(\tau,\tau') \right)
\, ,
\end{equation}
Using the scaling of the couplings with $N$ and by the same 
arguments developed for classical systems \cite{Pl,Crso2}, 
we have verified that all orders $n\geq 3$ in the series (\ref{series})
are suppressed in the thermodynamic limit. 

\subsection{A cavity interpretation}\label{meaning}

First of all, let us write the TAP equations in a way which allows 
one to clarify the physical meaning of the different terms:
\begin{eqnarray}\label{tap1mean}
\left. \frac{\delta (-\beta F)}{\delta  m_{i}(\tau )}
\right|_{\alpha =0}&=&h_{i}^{cav}(\tau )=
- \sum_{i_2 < \dots < i_p} J_{i, i_2,\dots,i_p} m_{i_2}(\tau) \dots
m_{i_p}(\tau)\\
&&-\frac{1}{\hbar }\int_0^{\beta \hbar } d\tau' 
\left[\frac{p(p-1)}{2} 
\left( Q(\tau,\tau') - C(\tau,\tau') \right)  Q^{p-2}(\tau,\tau')
\right] \, m_i(\tau') \; , 
\nonumber \\
\label{tap2mean}
\left. \frac{2}{N}\frac{\delta (-\beta F)}{\delta C(\tau, \tau'  )}
\right|_{\alpha =0}&=&G^{cav}(\tau,\tau ')=\frac{p}{2} \left[
Q^{p-1}(\tau,\tau') -  C^{p-1}(\tau,\tau') \right]
\; .
\end{eqnarray}
The solutions to these equations are expected to be 
time-translation invariant since we are developing  a description
of equilibrium and metastable properties. Therefore $h_{i}^{cav}$
is indeed independent of the imaginary time and $G^{cav}$ depends only
on the difference between $\tau $  and $\tau '$.

An understanding of the meaning of the quantum TAP equations
follows from the analysis of $F$ for $\alpha =0$. Indeed, by 
tracing out all the spins except $s_{i}$ in the partition function 
produces a single-site measure (for 
$s_{i}$) whose action reads:
\begin{equation}\label{singlesite}
-\frac{1}{\hbar}\int_{0}^{\beta \hbar }d\tau \left[
H_{k}(s_{i}(\tau ))+h^{cav}_{i}(\tau )s_{i}(\tau )\right]
-\frac{1}{2 \hbar^2} \int_0^{\beta\hbar} d\tau \int_0^{\beta\hbar} d\tau' \; 
s_i(\tau)G^{cav} (\tau,\tau') s_i(\tau') \;, 
\end{equation}
where $H_{k}(s_{i}(\tau ))$ is the kinetic energy for the spin $s_{i}$.
As a consequence the TAP solutions are the 
self-consistent relations that relate $G^{cav}(\tau ,\tau ')$
 and $h_{i}^{cav}(\tau )$ (which are functions of $C(\tau ,\tau ')$
and $m_{i}(\tau )$) to $C(\tau ,\tau ')$
and $m_{i}(\tau )$ obtained from the single-site action
(\ref{singlesite}).

Equations (\ref{tap1mean}) and (\ref{tap2mean}) show that
the action on the {\it i}th spin of the $N-1$ remaining ones 
reduces simply to $h_{i}^{cav}$ and $G^{cav}$. This implies 
that tracing out all the spins but the {\it i}th one
produces a Gaussian measure for 
the instantaneous magnetic fields $h_{i}(\tau)=-\sum_{i_{2}<\dots <i_{p}}
J_{i_{1},\dots ,i_{p}}s_{i_{1}}\cdots s_{i_{p}}$, whose mean and
connected two-point correlation
function equal respectively $h_{i}^{cav}$ and $G^{cav}(\tau -\tau ')$. 

\begin{figure}[bt]
\centerline{    \epsfysize=8cm
       \epsffile{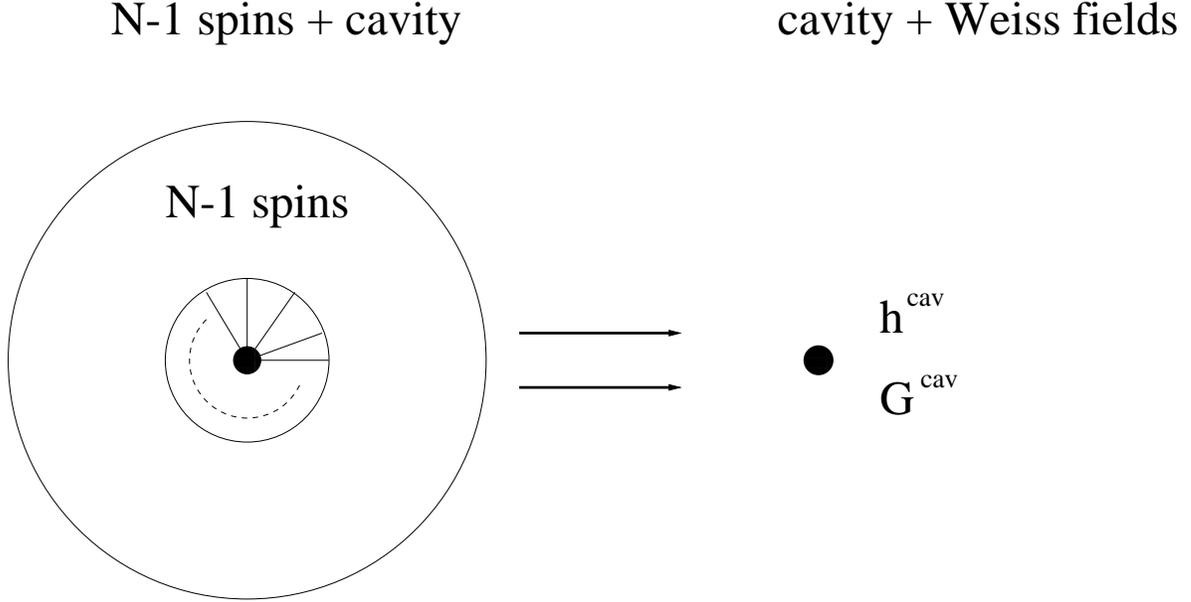}}
\caption{A schematic representation of the action of the $N-1$ spins
 on the cavity which, due to the infinite connectivity, simply reduces
to $h_{i}^{cav}$ and $G^{cav}$.}
\end{figure}

The expression of $h_{i}^{cav}$ and $G^{cav}(\tau -\tau ')$ can 
be justified within the cavity method \cite{Beyond}. 
Let us focus for simplicity on the $p=2$ case for which
\begin{equation}\label{cavp2}
h_{i}^{cav}=-\sum_{k}J_{i, k }
\langle s_{k}\rangle _{N-1}
\end{equation}
where $\langle\cdot \rangle _{N-1}$ represents the thermal average with respect to the
system with the {\it i}th site removed. 
$\langle s_k \rangle _{N-1}$ is not simply equal to $m_{k}$, which is 
the mean magnetization for the system of N spins. A correction term, 
first discovered by Onsager, appears:
\begin{equation}\label{onsager}
\langle s_k \rangle _{N-1}=m_{k}-\frac{1}{\hbar}
\int_{0}^{\beta \hbar}d\tau \frac{\delta m_{k}(\tau )}{h_{k}(\tau ')}
J_{i,k}m_{i}(\tau ')=m_{k}-\frac{1}{\hbar}\int_{0}^{\beta \hbar}
d\tau \left[ C(\tau) -Q\right]J_{i,k} m_{i} 
\end{equation} 
Plugging Eq.~(\ref{onsager}) into Eq.~(\ref{cavp2}) and using the scaling 
 of the couplings with $N$ one recovers the expression for 
$h_{i}^{cav}$ given in (\ref{tap1mean}) in the $p=2$ case. 
Whereas for $G^{cav}$ a similar computation \footnote{Indeed easier since 
there is no reaction term for $G^{cav}$ on a completely connected lattice.}
gives back the expression given in (\ref{tap2mean}).

Finally, we remark that 
the main difference between the classical and the quantum TAP approach
 is that in the latter the cavity interaction
consists not only in a cavity field but also in the ``Weiss function''
$G^{cav}(\tau -\tau ')$, which is a function of (imaginary) time. 
This already happens in the mean-field theory of quantum 
non-disordered systems \cite{Kotliar} for which local quantum fluctuations 
are taken into account exactly, whereas the spatial ones are frozen.
For disordered systems, even in the limit of infinite dimensions,
 one has to take into 
 account not only the local quantum fluctuations but also 
some spatial fluctuations: all the instantaneous magnetic fields 
 have the same variance but their averaged values fluctuate
 from site to site. 
 
\section{A continuous disordered quantum model}
\label{p-spin}

In this Section we apply the method of Section~\ref{sec1:formalism} 
to the study of the quantum spherical $p$ spin-glass
model. We derive and analyze the TAP free-energy density and the TAP equations
for the local magnetization and correlation function in
imaginary time. We relate these equations 
to the equation for the order parameter in the 
Matsubara replicated approach to equilibrium
and in the Schwinger-Keldysh approach to the 
non-equilibrium dynamics.
 
\subsection{The model and its TAP equations}
\label{model}

A model of a quantum particle with position ${\bf s}$ and momentum ${\bf p}$
that  moves on an $N$-dimensional random environment
is defined as 
\begin{equation}
H[{\bf p},{\bf s},J]= \frac{{\bf p}^2}{2M}  - \sum^{N}_{i_1<...<i_p} J_{i_1...i_p}
s_{i_1} ... s_{i_p}
\; .
\label{eq:action}
\end{equation}
A Lagrange multiplier $z$ enforces the 
averaged spherical constraint
\begin{equation}
{1\over N}\sum^{N}_{i=1} \langle s_i^2 \rangle = 1 
\; .
\label{eq:lm}
\end{equation}
The random interaction strengths $J_{i_1...i_p}$ are taken
with zero mean and variance defined in Eq.~(\ref{coupling}).
This model is a possible quantum extension of the 
spherical $p$ spin-glass model introduced in \cite{Crso1} and it is a particular realization of 
the class defined in (\ref{potential}) corresponding to space-coordinates $s_{i}$ constrained 
 to move on a N-dimensional sphere. 

The zero-th order term of the expansion (\ref{series}) can be readily computed for this model. 
By setting $\alpha=1$, rescaling time according to $\tau \to \tau\hbar/\tilde{J}$, and defining the ``quantum 
parameter'' $\Gamma \equiv \hbar^2/(\tilde{J}M)$ we obtain the following expression for the quantum TAP free-energy (\ref{eq:Gamma}) :
\begin{eqnarray}
& & -\beta F = 
 \frac{N}{2} \mbox{Tr} \, \mbox{ln} (C-Q) 
+ 
\frac{N}{2\Gamma} \mbox{Tr} \left( \frac{ \partial^2 C}{ \partial \tau^2 } \right)
+
\int_0^\beta d\tau \sum_{i_1< \dots< i_p} 
J_{i_1,\dots,i_p} m_{i_1}(\tau) \dots m_{i_p}(\tau)
\nonumber\\
& & \;\;\;\;\;\;
 +
\frac{N}{4} \int_0^\beta d\tau \int_0^\beta d\tau' 
\left( 
C^p(\tau,\tau') - Q^p(\tau,\tau') - p (C(\tau,\tau') - Q (\tau,\tau') ) Q^{p-1}(\tau,\tau') \right)
\nonumber\\
& &
\;\;\;\;\;\;
- \frac{N\beta}{2} \int_0^\beta d\tau 
z(\tau) \left( C(\tau,\tau)-1 \right)
\end{eqnarray}
The physical parameters $m_i(\tau)$ and $C(\tau,\tau')$ are fixed by the quantum TAP equations~(\ref{legendre})
\begin{eqnarray}\label{eq2}
& & h_i(\tau) =
 \sum_{i_2 < \dots < i_p} J_{i, i_2,\dots,i_p} m_{i_2}(\tau) \dots
m_{i_p}(\tau) \\
& & 
+ \int_0^\beta d\tau' 
\left[
-(C-Q)^{-1}(\tau,\tau')  + \frac{p(p-1)}{2} 
\left( Q(\tau,\tau') - C(\tau,\tau') \right)  Q^{p-2}(\tau,\tau')
\right] \, m_i(\tau')
\; ,\nonumber	
\\
& & 
z(\tau)\delta (\tau -\tau')  =
(C-Q)^{-1}(\tau,\tau') + \delta(\tau-\tau') \frac{1}{\Gamma} \frac{\partial^2}{\partial\tau^2} +
\frac{p}{2} \left[  C^{p-1}(\tau,\tau') -  Q^{p-1}(\tau,\tau') \right]
\label{eq1}
\; .\nonumber
\end{eqnarray}

Finally, setting $h_i(\tau)=0$ and using that at stationarity, $m_i(\tau) = m_i$,  $Q(\tau,\tau')= q_{\sc ea}$, $z(\tau)=z$
and $C(\tau,\tau')=C(\tau-\tau')$, the previous equations
are simplified to
\begin{eqnarray}
\frac{1}{\Gamma} \frac{\partial^2 C(\tau) }{\partial \tau^2} 
&=&
-\frac{p}{2}  \int_0^\beta d\tau'  
\left( C^{p-1}(\tau-\tau')-q_{\sc ea}^{p-1}\right) \left( C(\tau')-q_{\sc ea} \right)
\nonumber\\
& & 
+z \left(C(\tau)-q_{\sc ea}\right) -\delta(\tau)
\; ,
\label{eqC-Qstat}
\\ \hphantom{a}\nonumber \\
z m_i 
&=&
 \sum_{i_2< \dots < i_p} 
J_{i, i_2,\dots,i_p} m_{i_2}\dots m_{i_p} 
+
\nonumber \\
& & 
m_i \; \frac{p}{2} \int_0^\beta d\tau' \; 
\left( 
C^{p-1}(\tau') +(p-2) q_{\sc ea}^{p-1} 
-
(p-1) C(\tau') q_{\sc ea}^{p-2}
\right)  
\label{eqmstat}
\; .
\end{eqnarray} 

\subsection{Analysis of the quantum TAP equations}\label{analysis}

In the classical case the TAP equations admit 
a large number of solutions at low temperatures. In the following we shall 
show that this 
remains the case in a certain regime of $T$ and $\Gamma $. Furthermore
we shall classify them by their Edwards-Anderson parameters.
Finally, we shall exhibit several properties of the TAP solutions
valid at low temperatures.

\subsubsection{A simple equation on the Edwards-Anderson parameter}\label{eqqea}

Let us analyze in detail the equations for the local magnetizations, $m_{i}$. 
First of all
 we note that a simple equation that relates $q_{\sc ea}$ to the potential energy density
derives from Eq.~(\ref{eq2}). In fact, 
by multiplying Eq.~(\ref{eq2}) by $m_i/N$ and summing over $i=1,\dots N$
one obtains (for $h_{i}=0$)
\begin{equation}
0 = - \frac{q_{\sc ea}}{\tilde C(0)-\beta q_{\sc ea}} + 
\frac{p}{N}\sum_{i_1< \dots< i_p} J_{i_1\dots i_p} m_{i_1} 
\dots m_{i_p} - \frac{p(p-1)}{2} \left( \tilde C(0)-\beta q_{\sc ea}\right) q_{\sc ea}^{p-1}
\label{eqq1}
\end{equation}
where we introduced the discrete Fourier transform of the correlation
\begin{equation}
\tilde C(\omega) \equiv \int_0^\beta d\tau \; e^{i\omega\tau} C(\tau)
\; . 
\end{equation}
Following Kurchan {\it et al} \cite{Kupavi}, 
we introduce the {\it angular} variables $\sigma_i=m_i/\sqrt{q_{\sc ea}}$
and define the {\it angular potential energy density}
\begin{equation}
{\cal E}({\bf \sigma}) \equiv - \frac{1}{N} \sum_{i_1<\dots <i_p} J_{i_1\dots i_p} \, 
\sigma_{i_1} \dots \sigma_{i_p}
\; .
\end{equation}
For any fixed energy level ${\cal E}$, Eq.~(\ref{eqq1}) becomes a 
second-order polynomial equation for $q_{\sc ea}$; the solution
is determined by 
\begin{equation}
q_{\sc ea}^{p/2-1} (\tilde C(0)-\beta q_{\sc ea}) = z_\pm  =\frac{1}{p-1} 
\left( 
-{\cal E}(\sigma) \pm \sqrt{{\cal E}^2(\sigma)- {\cal E}_{\sc th}^2} 
\right)
\label{eqq1final}
\end{equation}
and ${\cal E}_{\sc th}$, called the threshold value,  is given by
\begin{equation}
{\cal E}_{\sc th} = -\sqrt{\frac{2(p-1)}{p}}
\; .
\end{equation}
The right-hand-side of Eq.~(\ref{eqq1final}) has to be real. This imposes the 
condition ${\cal E} \leq {\cal E}_{\sc th}$ since ${\cal E}$ is a negative quantity.

For each sign in Eq.~(\ref{eqq1final}), its
left-hand-side has a bell-shape
as a function of $q_{\sc ea}$. It vanishes at $q_{\sc ea}=0$ and $q_{\sc ea}=\tilde C(0)/\beta$
and attains its maximum at $q_{\sc ea}=(1-2/p)\tilde C(0)/\beta$.
Hence, at fixed values of ${\cal E}$ and $T$,  
Eq.~(\ref{eqq1final}) has none or   
{\it two} solutions, $q_{\sc ea}=q', q''$, with 
\begin{eqnarray}
0 \leq &q'& \leq  \frac{(1-2/p) \tilde C(0)}{\beta}
\; ,
\\
 \frac{(1-2/p) \tilde C(0)}{\beta} \leq &q''& \leq   \frac{\tilde C(0)}{\beta} 
\end{eqnarray}
(we assume, as expected, that $C(\tau )$ is positive for 
all $\tau $). 
In the classical case, the minus sign in Eq.~(\ref{eqq1final}) 
leads to a value of  $q_{\sc ea}$ that is a minimum of the TAP
free-energy for all ${\cal E} < {\cal E}_{\sc th}$. The Edwards-Anderson 
parameter determined in this way has the expected 
physical behavior~\cite{Crso2}. In Appendix A we 
show that in the quantum case one has to choose the minus sign 
in Eq.~(\ref{eqq1final}), too. Thus, $q_{\sc ea}$ is determined by 
\begin{equation}\label{eqq1finalnew}
q_{\sc ea}^{p/2-1} (\tilde C(0)-\beta q_{\sc ea}) = \frac{1}{p-1} 
\left( -{\cal E}(\sigma) - \sqrt{{\cal E}^2(\sigma)- {\cal E}_{\sc th}^2} \right)
\; .
\end{equation}
This equation still has two solutions. It can be proven that the solution 
with the larger absolute value of $q_{\sc ea}$ has the correct physical 
properties. In particular, it is connected to the classical solution, and 
it is then the solution to be kept. Thus there is a one to one
correspondence
between $q_{\sc ea}$ and ${\cal E}$.

It is of particular interest, as we shall show below, 
the {\it threshold solution} ${\cal E}={\cal E}_{\sc th}$.
In this case the equation for $q_{\sc ea}$ becomes
\begin{equation}\label{qea-eq}
1 = \frac{p(p-1)}{2} \; (\tilde C(0) - \beta q_{\sc ea})^2 q_{\sc ea}^{p-2}
\; .
\end{equation}

Note that this equation coincides with the one found with 
the Matsubara formalism using the 
marginality condition to fix the block size $x$ in the replica matrix \cite{Cugrsa}.
Furthermore, it coincides with the equation for the dynamic value of the Edwards-Anderson
parameter $q_{\sc ea} \equiv \lim_{t\to\infty} \lim_{t_w\to\infty} C(t+t_w,t_w)$ 
obtained from the study of the real-time dynamics of the quantum model 
evolving in contact with an Ohmic quantum environment \cite{Culo},
when one takes first the thermodynamic limit, next the long-time 
limit of the system's dynamics in contact with the environment and, finally, 
the strength of the coupling to the environment to zero. 
The relationship between TAP, Matsubara and dynamical approach will be discussed in Section \ref{reltapmatdyn}.  

\subsubsection{Multiplicity of TAP solutions}\label{subsec:complexity}

The equations (\ref{eqq1final}) reveal an interesting structure of the
 TAP equations (\ref{eqC-Qstat}) and (\ref{eqmstat}). For a given value of the 
 angular potential energy ${\cal E}$ the TAP equations decouple
 in two different sets: Eqs.~(\ref{eqC-Qstat}) and (\ref{eqq1final}),
with the spherical condition on $C$,
determine the correlation function, the spherical parameter and
 the Edwards-Anderson parameter; whereas Eqs.~(\ref{eqmstat}) 
determine the angular variables only. They read
\begin{eqnarray}\label{angular}
\mu q_{\sc ea}^{1-p/2} \sigma_i &=& -p  {\cal E}({\bf \sigma})\sigma_i=
p \sum_{i_2 < \dots < i_p} 
J_{i, i_2,\dots,i_p} \sigma_{i_2}\dots \sigma_{i_p}
\; ,
\\
\mu & \equiv &z- \frac{p(p-2)\beta}{2} q_{\sc ea}^{p-1} + 
\frac{p(p-1)}{2} \tilde C(0) q^{p-2} - \tilde \Sigma(0) 
\; ,
\label{eq-mu}
\end{eqnarray}
where we have defined
\begin{equation}
\tilde \Sigma(0) \equiv \frac{p}{2} \int_0^\beta C^{p-1}(\tau) 
\; .
\end{equation}
For a given value of the angular potential energy ${\cal E}$ 
Eqs.~(\ref{angular}) allow one to determine the 
angular part of the TAP solutions. 

\begin{figure}[bt]
\centerline{    \epsfysize=8cm
       \epsffile{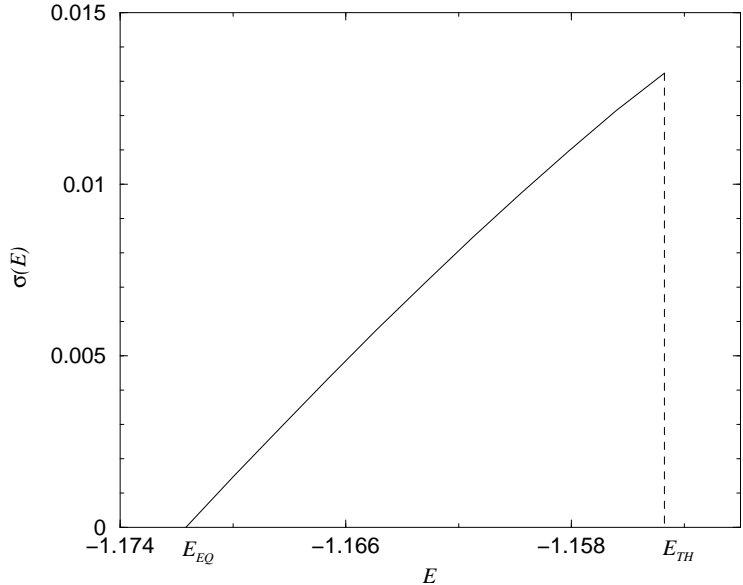}}
\caption{The complexity $\sigma({\cal E})$ 
as a function of ${\cal E}$ in the interval 
${\cal E}_{\sc eq}<{\cal E}<{\cal E}_{\sc th}$ for $p=3$.
\label{comp.fig}}
\end{figure}

In general for a given value of ${\cal E}$, Eqs.~(\ref{eqC-Qstat}) 
and (\ref{eqq1final}) determine  the correlation function, 
the spherical parameter and the Edwards-Anderson parameter {\it in a unique way}.
(An exception to 
 this rule are the paramagnetic solutions which however 
do not correspond to any ${\cal  E}$.) As a consequence, the multiplicity of 
TAP solutions is entirely due to Eq.~(\ref{angular})
 which, for certain values of ${\cal  E}$, can admit an exponential (in $N$)
 number of solutions ${\cal N}({\cal E})$. The complexity 
as a function of ${\cal E}$ is then defined as 
\begin{equation}
\sigma({\cal E}) \equiv \lim_{N\to\infty} \frac{1}{N} 
\overline{\ln( {\cal N}({\cal E}) )}
\; ,
\end{equation}
Equation~(\ref{angular}) already appears at the classical level. 
The complexity has been computed by Crisanti and
Sommers \cite{Crso2} and Cavagna {\it et al.} \cite{Cagipa1} with the following 
result. There are typically no 
solutions for ${\cal E}<{\cal E}_{\sc eq}$, whereas for ${\cal E}_{\sc eq}
<{\cal E}<{\cal E}_{\sc th}$ the complexity reads
\begin{eqnarray}\label{complexity}
\sigma({\cal
E})&=&\frac{1}{2}\left(1+\ln\left(\frac{p}{2}\right)\right)
-
{\cal E}^2+
\left(\frac{{\cal E}- \sqrt{{\cal E}^2-{\cal E}_{\sc th}
        ^2}}{\sqrt{2}{\cal E}_{\sc th}}\right)^2 + 
\ln\left(-{\cal E}- \sqrt{{\cal E}^2-{\cal E}_{\sc th} ^2}\right)\nonumber
\\
&&\qquad \qquad\qquad \qquad\qquad \qquad {\mbox {for}}
\quad \quad 
{\cal E}_{\sc eq}<{\cal E}<{\cal E}_{\sc th}
\end{eqnarray}
where ${\cal E}_{\sc eq}$ is the value at which $\sigma({\cal E})$ vanishes.
A plot of this function is traced in Fig.~\ref{comp.fig} for $p=3$.
 
\subsubsection{A low temperature and low $\Gamma $ approximation}
\label{approximate}

In the classical case the TAP equations separate  
in two sets: 
$N$ equations for the angular variables and one for the Edwards-Anderson parameter. 
The former
 admit an exponential number of solution and are studied from a statistical
 point of view (one computes the number of solution, and the typical 
 properties of
 solution corresponding to a given ${\cal {E}}$), whereas the latter
 can be easily solved. In the quantum case the analysis of the equations  
for the angular variables is identical to the one used for classical systems.
The analog of the equation that determines $q_{\sc ea}$ becomes now 
a differential equation for $C(\tau)$
 that has to be studied numerically. This differential equation can be mapped 
exactly onto the ones analysed with the replica method, as we shall show in 
Section (\ref{reltapmatdyn}), and its numerical solution can be found
 in \cite{Cugrsa}. In this section we perform a low-temperature 
and low $\Gamma $ approximation, also discussed in \cite{Cugrsa}, 
that allows one to obtain some 
qualitative results that 
remain valid for the exact solution.

At low temperature and low $\Gamma $, the extension of the imaginary 
time-interval diverges $[0,\beta\to\infty]$ and the periodic 
correlation $C(\tau)$ is expected to have a rapid decay, over a 
short time-interval, from $1$ to its ``asymptotic'' value,
say, at $\tau=\beta/2$. Moreover the ``regular'' 
part of the correlation, that we define as \cite{Cugrsa}
\begin{equation}
q_{\sc reg}(\tau)  \equiv C(\tau) - q_{\sc ea} 
\end{equation}
can be assumed to be small. Therefore we can expand the TAP free-energy 
 in powers of $q_{\sc reg}(\tau)$. Up to terms of the order of 
$q_{\sc reg}(\tau)^{3}$ we obtain
\begin{eqnarray}\label{tapapprox}
& & -\frac{\beta F}{N} = 
 \frac{1}{2} \mbox{Tr} \, \mbox{ln} (q_{\sc reg}(\tau)) 
+ 
\frac{1}{2\Gamma} \mbox{Tr} 
\left( \frac{ \partial^2 q_{\sc reg}(\tau)}{ \partial \tau^2 } \right)
+
\frac{\beta }{N}  \sum_{i_1< \dots< i_p} 
J_{i_1,\dots,i_p} m_{i_1} \dots m_{i_p}
\nonumber\\
& & \;\;\;\;\;\;
 +
\beta \frac{p(p-1)}{4}q_{\sc ea}^{p-2} \int_0^\beta d\tau q_{\sc reg}^{2}(\tau )
- \frac{\beta}{2}  
z  \left(q_{\sc reg}(0)+q_{\sc ea}-1 \right)
\; ,
\end{eqnarray}
where we have focused on the space of time translation 
invariant (TTI) functions (since the TAP solutions are TTI this 
 does not imply a loss of generality).

Within this approximation the TAP equations become quadratic in Fourier space, 
\begin{equation}
1- \left( \frac{w^2_k}{\Gamma} + z \right) \tilde q_{\sc reg}(\omega_k) + 
\frac{p(p-1)}{2} \, q_{\sc ea}^{p-2} \, \tilde q_{\sc reg}^2(\omega_k) =0
\; ,
\end{equation}
 and yield
\begin{equation}\label{qapprox}
\tilde q_{\sc reg}(\omega_k) = 
\frac{z + \omega_k^2/\Gamma \pm \sqrt{(z + \omega_k^2/\Gamma)^2
- 2 p (p-1) q_{\sc ea}^{p-2}}}{p (p-1) q_{\sc ea}^{p-2}}
\; .
\end{equation}
By taking $\omega_k=0$ and comparing to Eq.~(\ref{eqq1finalnew}) one obtains
\begin{equation}
{\cal E} = - \frac{z q_{\sc ea}^{1-p/2}}{p}
\; .
\end{equation}
The spherical constraint reads
\begin{equation}
1-q_{\sc ea} = \frac{1}{\beta} \sum_k \tilde q_{\sc reg}(\omega_k) = \int_0^\infty \frac{d\omega}{\pi} \; \chi''(\omega) \; \coth\left(\frac{\beta\omega}{2}\right)
\label{eq:constaint}
\end{equation}
where 
\begin{equation}
\chi''(\omega) \equiv \mbox{Im} \; \tilde q_{\sc reg}(\omega_k=-i\omega)
=
\frac{q_{\sc ea}^{1-p/2}}{p-1} \sqrt{{\cal E}^2_{\sc th} - 
\left({\cal E}+\frac{\omega^2 q_{\sc ea}^{1-p/2}}{p\Gamma} \right)^2 }
\; .
\label{eq:Gamma-q1}
\end{equation}
The integral in Eq.~(\ref{eq:constaint}) has to be taken on the interval 
$\omega \in [\omega_-,\omega_+]$ such that the square root is real. 

In the low temperature limit, we approximate $\coth(\beta\omega/2)\sim 1$ and, 
by changing variables in the integral,  we obtain 
\begin{equation}
\Gamma I^2({\cal E}, p) = \frac{\pi^2 (p-1)^2}{p} (1-q_{\sc ea})^2 q_{\sc ea}^{(p-2)/2}
\end{equation}
with 
\begin{equation}
I({\cal E}, p)= 
2\int_{\sqrt{-{\cal E}+{\cal E}_{\sc th}}/2}^{\sqrt{-{\cal E}-{\cal E}_{\sc th}}} 
dx \, \sqrt{{\cal E}^2_{\sc th} - \left({\cal E}+ x^2\right)^2 }
\; .
\end{equation}
Equation (\ref{eq:Gamma-q1}) yields a relation between $\Gamma$, 
$q_{\sc ea}$ and ${\cal E}$ of the form 
\begin{equation}
\Gamma I^2({\cal E},p) = \mbox{ct} (1-q_{\sc ea})^2 q_{\sc ea}^{(p-2)/2}
\; ,
\end{equation}
with $\mbox{ct}$ a numerical constant.
For each ${\cal E}$, there is a 
solution with a  physically meaningful value of $q_{\sc ea}$ that is close to $1$,  
until  reaching a critical $\Gamma_{\sc max}({\cal E})$. 
This value tells us when the TAP solutions associated to 
${\cal E}$ disappear. It can be easily proven that 
$I({\cal E},p)$ is a growing function of  ${\cal E}$; hence, 
$\Gamma_{\sc max}({\cal E})$ is a decreasing function of ${\cal E}$. 
This implies that the TAP solutions that are at the threshold level 
disappear first than those that are at lower values of ${\cal E}$. 
This is again similar to the dependence of the classical TAP solutions with 
temperature \cite{Kupavi}: the solutions corresponding to the threshold 
level disappear at a lower temperature  than the ones corresponding 
to the equilibrium level $T_{\sc max}({\cal E}_{\sc th})<T_{\sc max}({\cal E}_{\sc eq})$
and, more generally,  $T_{\sc max}({\cal E}_1)<T_{\sc max}({\cal E}_2)$
if ${\cal E}_1 > {\cal E}_2$.

The low frequency behavior of the spectral density 
$\chi''(\omega) \equiv \mbox{Im} \; \tilde q_{\sc reg}(\omega_k=-i\omega)$
of the threshold states is gapless,  
\begin{equation}\label{chinogap}
\chi''(\omega)\sim \omega \qquad {\mbox{for}} \quad \omega \rightarrow 0^{+}
\; ,
\end{equation}
whereas all the other states (${\cal {E}}<{\cal {E}}_{\sc th}$) have 
a gap $\Delta $ in their excitation spectrum, 
\begin{equation}\label{chinogap2}
\chi''(\omega)\sim \sqrt{\omega-\Delta } \qquad {\mbox{for}} 
\quad \omega \rightarrow \Delta^{+}
\; .
\end{equation}

Furthermore we have studied the dependence
 of the free-energy on $\cal {E} $ and $\Gamma $ in the low temperature limit. Plugging the solution
(\ref{qapprox}) into (\ref{tapapprox}) we find, after a tedious computation,
\begin{equation}\label{freeenergyapprox}
-\frac{\beta F}{N} =-\int \frac{d\omega}{\pi \Gamma}
\ln \left(2\sinh \left(\frac{\beta \omega }{2}\right)\right)
\; \omega \chi''(\omega)+\frac{\beta q_{\sc ea}^{p/2}}{2} 
\left(p-2-\frac{p}{q_{\sc ea}} \right) {\cal {E}}
\end{equation}
where $q_{\sc ea}$ satisfies Eq.~(\ref{eqq1finalnew}).
This expression allows one to 
study the evolution of the free-energy of the TAP states as a function
 of $\Gamma $. We have found that
if one knows a TAP solution at zero temperature and zero $\Gamma $ 
one can  follow it continuously in $\Gamma$. As in the classical case
TAP solutions do not cross, merge nor divide in this model.

\begin{figure}[bt]
\centerline{    \epsfysize=8cm
       \epsffile{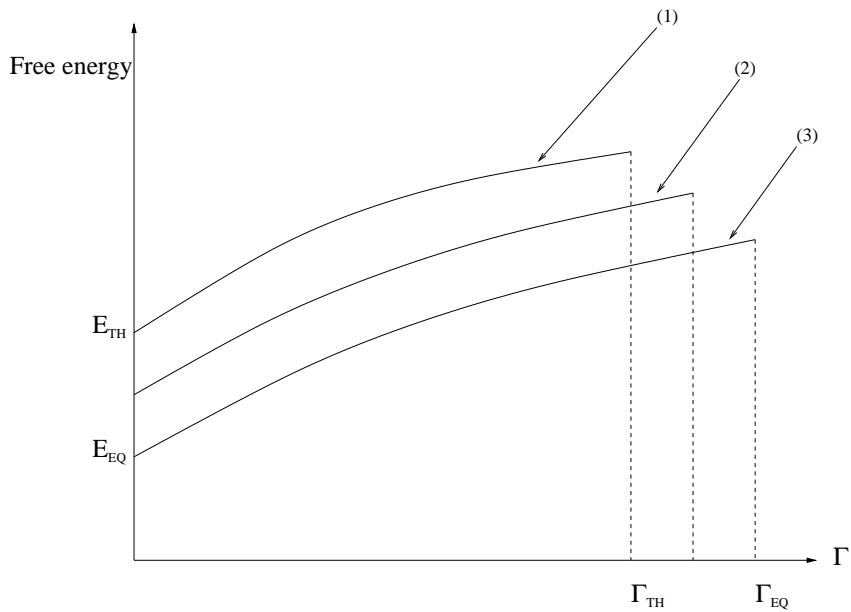}}
\caption{TAP free-energy versus $\Gamma $ at zero temperature. The curve
 (1) corresponds to threshold states which are the first ones to disappear 
 at $\Gamma _{\sc th}$; the curve (3) corresponds to the states with the lowest
 free-energy which are the last ones to disappear at $\Gamma _{\sc rsb}$ and the
 curve (2) corresponds to an intermediate state.
\label{fgamma.fig}}
\end{figure}

Figure~\ref{fgamma.fig} summarizes these results in a schematic way. 
Finally, note the special role of the threshold states, which are gapless
 (contrary to the others), the first ones to disappears and the ones
 with highest free-energy density.

\subsubsection{Stability of TAP states}
\label{properties}

In the classical case one can check the stability of TAP states.
 In the quantum case this is
 a difficult task that has to be performed numerically. In the following
 we shall limit ourself to prove that the TAP solutions characterized
 by ${\cal {E}}={\cal {E}}_{\sc th}$ (threshold states) are characterized by zero modes
and are hence marginally stable. 
We expect that a complete
 stability analysis will confirm that the TAP states
 characterized by ${\cal {E}}<{\cal {E}}_{\sc th}$ are stable.

Let us focus on the reduced free-energy Hessian $\partial^{2}F
/\partial m_{i}(\tau )\partial m_{j}(\tau' )$ evaluated in a TAP solution
 $\{m_{i}^{\alpha } \}$.
 This matrix depends on $\tau, \tau '$ only through their difference.
 Therefore it is diagonal in Fourier
 space. Focusing on zero frequency, the original problem reduces to the 
 diagonalization of the following matrix:
\begin{equation}
\label{hessian}
A_{i,j}=-\sum_{i_{3}<\dots <i_{p}}J_{i,j,i_{3},\dots ,i_{p}}m_{i_{3}}^{\alpha }
\cdots m_{i_{p}}^{\alpha }-p{\cal {E}}q_{\sc ea}^{p/2-1}\delta _{i,j}
\; ,
\end{equation}
where $q_{\sc ea}=\sum_{i}(m_{i}^{\alpha })^{2}/N$.
The density of eigenvalues of $\mathbf{A}$ has been computed in \cite{replicon} and,
 except for the isolated eigenvalue corresponding to the eigenvector 
$m_{i}^{\alpha }$, it is a semicircular law centered in 
$-p{\cal {E}}q_{\sc ea}^{p/2-1}$ with width $-p{\cal {E}}_{\sc th}q_{\sc ea}^{p/2-1}$. 
Consequently, threshold states are characterized
by a vanishing fraction of zero modes.

\subsubsection{The classical limit}\label{classical}

The classical limit of Eqs.~(\ref{eqC-Qstat}) and (\ref{eqmstat})  yields
 the classical TAP equations computed by Kurchan {\it et al.} \cite{Kupavi}.
In fact, in the classical limit, $C(\tau)=1$ and the parameter $z$
is fixed  by integrating Eq.~(\ref{eqC-Qstat})
between $0^+$ and $\beta^+$. This yields
\begin{equation}
z = \frac{1}{\beta(1-q)} + \frac{p\beta}{2} \left( 1- q^{p-1}\right
)
\; .
\end{equation}
By inserting this value of $z$ in Eq.~(\ref{eqmstat})
we obtain 
\begin{equation}
\left( \frac{1}{\beta(1-q)} + \frac{p(p-1)\beta}{2} (1-q) q^{p-2}
\right) m_i = 
p \sum_{i_2 < \dots < i_p} 
J_{i, i_2,\dots,i_p} m_{i_2} \dots m_{i_p}
\; ,
\label{eq-sigma_class}
\end{equation}
that coincide with the classical TAP equations for the local magnetizations.
The equation that fixes $q_{\sc ea}$ as a function of ${\cal E}$
in the classical limit is simply obtained from 
Eq.~(\ref{eqq1final}) by setting $\tilde C(0)=1$. 

\subsection{Relation between TAP, Matsubara and dynamic approaches}
\label{reltapmatdyn}

In Section \ref{introtap} we have recalled  
the relationship between TAP, replica and dynamical approaches 
in the classical case. In this Subsection 
we show how these connections are generalized to quantum systems.

\subsubsection{TAP-Matsubara}
 
Via the replica analysis in the Matsubara imaginary-time
framework  and within a 1step RSB Ansatz, the order parameter is 
the $n\times n$ matrix $Q_{ab}$  which is fully described by:
$n$ identical diagonal elements $q_d(\tau)$ that depend on the 
the imaginary time $\tau$, $n(x^2-1)$ constant elements $q_{\sc ea}$ that occupy the 
$x\times x$ blocks around the diagonal, the remaining $n^2-n(x^2-1)$ elements 
$q_0$ that in the absence of an external field are identically zero. 
In the $n\to 0$ limit, the three parameters $q_d(\tau)$, $q_{\sc ea}$ and $x$, 
together with the value of the Lagrange multiplier that enforces the 
averaged spherical constraint
determine the full solution of the problem~\cite{Cugrsa}. 

The connection between TAP and the Matsubara approaches 
is obtained by identifying the  Edwards-Anderson parameters $q_{\sc ea}$ in the
two approaches, $C(\tau)$ with the 
$\tau$-dependent diagonal parameter $q_d(\tau)$ in $Q_{ab}$, 
and the Lagrange multipliers. In particular
 we have shown that
subtracting the equation obtained for $a\neq b$ from the one corresponding to
 $a=b$ one obtains Eq.~(\ref{eqC-Qstat}). 
In the Matsubara approach one has another equation for 
$q_{\sc ea}$ in which $x$ acts as an external parameter.
Therefore by fixing the value of the breakpoint one fixes the value of 
$q_{\sc ea}$. As in the classical case
two different recipes to fix the breaking point parameter $x$, namely optimization and 
the marginality condition, lead to the static and dynamic transitions, respectively. 
In the TAP approach the role of $x$ is played by ${\cal  E}$ that
enters the
equation for 
$q_{\sc ea}$ as a parameter. 
We have found that 
the relationship between $x$ and $\cal E$ is encoded in 
\begin{equation}\label{mE}
\beta x=\frac{\partial\sigma(\beta, f) }{\partial f}
\; ,
\end{equation}
where $\sigma $ is the complexity defined in (\ref{complexity0}), see
Appendix B. 
This suggests that
in the a quantum problem the relationship (\ref{reltaprep}) is generalized 
to
\begin{eqnarray}\label{reltaprepquantum}
-\lim_{N\rightarrow \infty }\frac{1}{\beta N}\sum_{\alpha }e^{-\beta x Nf_{\alpha }}
&=&-\lim_{N\rightarrow \infty }\frac{1}{\beta N}\ln \int df e^{N(-\beta xf+\sigma (\beta,f))}
\\
&=&
x{\mbox {Extr}}_{q_{\sc ea},q_{d}(\tau )}f_{\sc rep}(q_{\sc ea},q_{d}(\tau );x,\beta,\Gamma )\nonumber
\end{eqnarray}
Using Eq.~(\ref{mE}) we have found that the Matsubara equations for $q_{\sc ea}$ 
 and $q_{d}(\tau )$ and the TAP equations for $q_{\sc ea}$ and $C(\tau )$ coincide.
 For instance, the TAP equations
 for the highest TAP states (threshold states) and the lowest TAP states
 coincide with the ones obtained in the Matsubara approach
 by using  the marginality condition and the extremization with respect to $x$,
 respectively.   
Moreover, as another confirmation of Eq.~(\ref{reltaprepquantum}) we have verified
 that the free-energy obtained from the Matsubara computation \cite{Cugrsa}
equals
 the one obtained in the TAP approach for all 
values of $\beta $ and $\Gamma $. In 
other words, we have checked that
\begin{equation}\label{check}
-\beta F=\ln \sum_{\alpha }e^{-\beta F_{\alpha }}
\; .
\end{equation}  
As a consequence the phase diagram that follows from the TAP approach coincides
with the one obtained in \cite{Cugrsa}.    

\subsubsection{TAP-out of equilibrium dynamics}. 

The study of the real-time dynamics of the p-spin quantum model 
evolving in contact with an Ohmic quantum environment has been
 performed in \cite{Culo}. The dynamical behavior is characterized 
 by two regimes. At high temperature and high $\Gamma $ the system 
 equilibrates in the paramagnetic state via an equilibrium dynamics.
Whereas at low temperature and low $\Gamma $ the systems ages and remains out of equilibrium also at infinite times. As in the classical case we have found 
that the long-time out of equilibrium dynamics is dominated by the threshold states.
This is proven by the fact that the equations for $q_{\sc ea}$ and $C(\tau)$ within the TAP approach coincide with the ones
obtained from the study of the real-time dynamics,
when the following limits are taken in its precise order: 
$\lim_{\gamma\to 0} \lim_{t\to\infty} \lim_{N\to \infty}$.
In order words, when one takes first the thermodynamic limit, next the 
long-time limit of the system's dynamics in contact with the environment 
and, finally, the strength of the coupling to the environment $\gamma$ to zero.
Notice that this equivalence holds for the paramagnetic states also.
Finally, we have shown that the relationship between the effective 
 temperature \cite{Cukupe} arising in the asymptotic out of equilibrium
 regime \cite{Culo} and the complexity is, as classically, 
\begin{equation}
\label{Teff}
\frac{1}{T_{\sc eff}}=\left. \frac{\partial\sigma(\beta, f) }
{\partial f}\right|_{f=f_{\sc th}}
\; .
\end{equation}

Note that the connection between TAP and real-time dynamics
is done by identification of several equations. A more precise 
analysis, along the lines of \cite{Bi}, should prove the full 
equivalence of the two methods.

\section{The phase diagram of discontinuous glassy systems}\label{transition}

In this section we present some general arguments that allow one to
predict the phase diagram of discontinuous glassy systems. Since 
the free-energy landscape plays a key role, 
we expect these results to have a certain degree of universality and
to apply to this entire class of disordered systems.

\subsection{The static transition}

Let us focus on two limiting regions of the phase diagram: around the classical phase transition
 ($\Gamma =0$) and around the quantum phase transition ($T=0$).

\begin{figure}[bt]
\centerline{    \epsfysize=8cm
       \epsffile{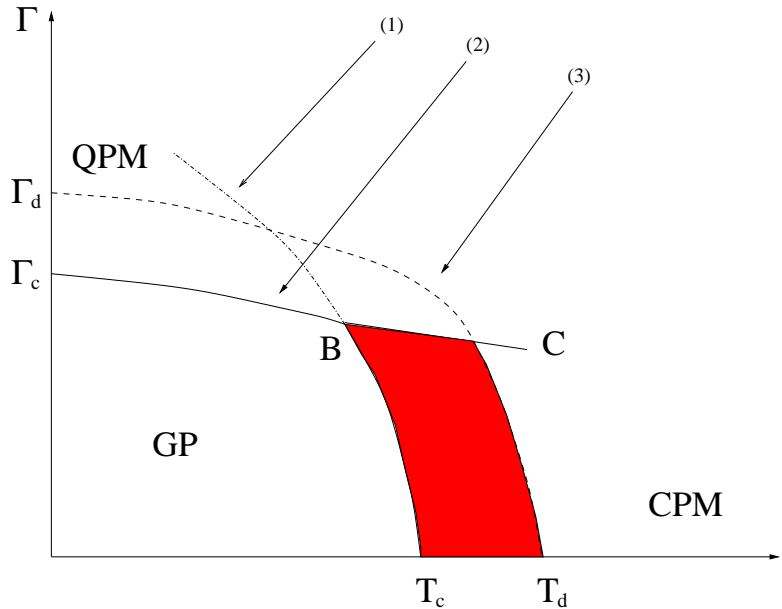}}
\caption{A schematic representation of the phase diagram which is
expected to be generic for systems having a discontinuous transition 
in the classical limit. 
The line $(1)$, from $(T_c,0)$ to the point B, represents the static transition between 
the classical paramagnet (CPM) and the classy phase (GP). 
The region in red between the lines $(1)$ and $(3)$ 
is the phase in which the CPM is fractured into an exponential number of TAP states. 
The line (2) from the point B to $(\Gamma_c,0)$, signals the static transiton 
between the quantum paramagnet (QPM) and the GP.
The line $(3)$ indicates the dynamic transition $T_{d}$ as a function of $\Gamma$.
\label{diag.fig}}
\end{figure}

In the former case the physics is well known and it is reviewed in 
Section \ref{introtap}. 
The effect of switching on weak quantum fluctuations consists only 
in a weak variation of the complexity $\sigma $. 
For this reason 
 the effect of quantum fluctuations reduces simply to a variation 
 of the thermodynamic ($T_{s}$) and the dynamic ($T_{d}$) transition 
temperatures (respectively lines $(1)$ and $(3)$ in Fig.~\ref{diag.fig}).

At zero temperature and low $\Gamma $ the system is in 
the glassy phase (GP), whereas at very high $\Gamma $ quantum fluctuations
destroy the glassy phase and the system is a quantum 
paramagnet (QPM). As a consequence 
one expects that a quantum phase transition should divide these two regimes
 at a certain value $\Gamma_{c}$.

At zero temperature the complexity is expected to remain a smooth function of 
the free-energy density\footnote{For instance the 
complexity does not blow up for $T\rightarrow 0$ since it is bounded by the 
logarithm of the number of energy minima divided by $N$, which is a finite quantity 
independent of temperature.}. Consequently,
equation~(\ref{eq:weighted3}) implies that 
the sum over the exponential number of glassy states is always dominated
by the lowest ones in free-energy since the $\beta f$ term in the
exponential largely dominates in this limit (this is different from
the classical problem in which other states dominate 
between $T_s$ and $T_d$). At zero temperature these are 
the states with lower angular potential energy, {\it i.e.} with
${\cal E}={\cal E}_{\sc eq}$. Now, from Fig.~\ref{comp.fig} we conclude that
$\sigma({\cal E}_{\sc eq})=0$ at zero temperature and for all $\Gamma$.
For this reason the mechanism behind the 
transition must be totally different from the classical one.    
 The transition cannot be related a configurational 
 entropy that vanishes when approaching $\Gamma_c$ from above 
(``entropy crisis'') since this quantity is always zero 
 at zero temperature.

Indeed, according to Eq.~(\ref{mE}), if we assume that $\partial
\sigma(\beta,f)/\partial f < +\infty$ when $T\to 0$, 
then $x\to 0$ for all $\Gamma$ in the glassy phase.
In the paramagnetic phase instead $x=1$. Thus, 
$x$ must jump at the transition. If the Edwards-Anderson 
parameter also jumps at $\Gamma_c$, the susceptibility is discontinuous, and 
the transition is of first order thermodynamically. As in the previous case
 the effect of switching on thermal fluctuations reduces simply, for low $T$, 
to a variation of $\Gamma _{c}$
 (line (2) in Fig.~\ref{diag.fig}).

Another hint on the difference between 
 classical and quantum phase transition can be gained by a technical 
 remark. It is well known that the paramagnetic solution 
of the classical problem remains stable  in the low temperature 
phase. This is a spurious solution of the mean field equations which 
 has to be discarded in the analysis of the low temperature regime. 
In the quantum case, one also expects 
 to find a spurious paramagnetic solution, which is the continuation
 of the classical paramagnet to low temperatures. 
 This solution exists to the left of line $(1)$ in Fig.~\ref{diag.fig},
 consequently one expects 
coexistence of two paramagnetic solutions: a physical one which is the 
 continuation of the quantum paramagnet valid at low temperatures and 
 high $\Gamma$ and a spurious one which is the continuation of the classical
 paramagnet. 

In the classical case the transition is of second order even if 
 the order parameter jumps discontinuously. This peculiar behavior is due 
to the fact that near the transition the paramagnetic state is fractured
 into an exponential number of states which
 continuously become  the ones responsible for the glassy phase at low 
temperature. This is not possible at zero temperature 
(the quantum paramagnet is not formed by a collection of glassy states)
and therefore it is reasonable to expect a quantum first 
 order phase transition between the glass phase and the quantum 
paramagnet.\footnote{Note however that other scenarios are possible. For example 
 the number of glassy states could diminish when $\Gamma$ increases and vanish 
 exactly at $\Gamma_{c}$. In this case the glassy states could be grown 
up continuously 
 from the quantum paramagnet.}

Finally, note that $F_{\sc qpm}=F_{\sc gp}=F_{\sc cpm}$ on the point B.
We then expect that a first order transition line 
separating the QPM from the CPM starts at this point. This line
should end on a point $C$ given that for very large value of
 $\Gamma $ and $T$
 the quantum and thermal fluctuations are so strong that the system
 becomes non interacting and in this case only one paramagnetic phase 
 exists. 
In the analysis of the $p=3$ spherical spin-glass model \cite{Cugrsa} 
the line $BC$ has not been found. We conjecture that in this case
the line $BC$ is so short that it is very difficult to find numerically. 
Furthermore, within the accuracy of the algorithm, the dynamic and static 
critical lines collapse at the point B. 
In the quantum model studied in \cite{Niri} instead this line has been detected
and it was demonstrated in this paper that its length  increases with $p$.

\subsection{The dynamic transition}

Now that the equilibrium phase diagram is completely predicted from a 
qualitatively point of view, we can focus on the non-equilibrium regime.
As noted previously, low quantum fluctuations simply change the values 
 of $T_{d}$ but do not change qualitatively the dynamic transition
  which remains
 second order in the sense that the asymptotic 
energy is continuous across the 
transition, but its derivative is not. This remains 
 true until the line $(2)$ reaches the line $BC$. After this point 
 the dynamic transition between the the quantum paramagnet and the 
  threshold states becomes first-order, i.e. the asymptotic energy is 
 not continuous across the transition. This, of course, is very
difficult to see numerically since the discontinuity has a very small 
value. 

\subsection{Summary}

In summary, through some general arguments based on the TAP approach
we have predicted a phase diagram that 
should
 have a certain degree of universality since its form is determined 
by the qualitative form of the free-energy landscape. 
Indeed, not only the quantum $p$ spin spherical model exhibits the phase diagram 
displayed in Fig.~\ref{diag.fig} but some other classical 
and quantum models share exactly 
the organization of phases and transitions \cite{Gash,Mosh,Niri,Opper}. 

\section{Conclusion}\label{conclu}

In this paper we have derived TAP equations for a large class of mean
 field disordered quantum system. Moreover we have applied the TAP approach
 to the quantum version of the spherical $p$ spin model. The study of this system,
 whose  real-time dynamics and statics have been analyzed 
in \cite{Culo,Cugrsa}, 
 has  furnished an ideal benchmark to 
 generalize to the quantum case several concepts 
developed for classical disordered systems. Armed with this knowledge,
 founded on the study of the free-energy landscape,
 we have shown that the same phase diagram, presented in
Fig.~\ref{diag.fig},
naturally emerges in  a large class of quantum disordered systems,
those having a classical discontinuous transition.

Whether other models like the SK model in a transverse field, 
or its soft spin version studied in \cite{Chamon}, also have such 
crossover in the transition from the disordered  to the ordered phase
is an issue that deserves revision. For the moment, no study of models
with classical continuous transition has shown this feature. However, 
it might have been masked by the methods used in previous studies.
The soft SK model might be the easiest example where to 
answer this equation via, {\it e.g.}, a careful application of the replicated 
Matsubara approach. 
 
We would like to stress that the TAP approach furnishes 
an alternative and more transparent route to 
replicas which has also the advantage of showing explicitly  
the weakness of the mean field description. Let us cite one example.
The marginality prescription in the replica approach becomes 
the more transparent statement that the non-equilibrium dynamics
is dominated by the TAP states which are marginally stable, {\it i.e.}
the flatness of the free-energy  landscape around these states 
is responsible for aging. Concerning one of the 
 weaknesses of the mean field description
 we would like to underline that the enormous number
 of pure states (with different free-energy densities) found 
for mean field models cannot persist in finite dimensions and the majority
 of them should become {\it metastable} states. How this changes the mean
 field scenario is an active domain of research for classical systems
\cite{Biku}.

We remark that interesting continuations of our work 
 concern, on the one hand, the application of the TAP approach to 
different quantum mean-field models \cite{quantum_mf,rotors,Cesare} 
and, on the other hand,
 the generalization of the static quantum TAP approach to real-time dynamics
 (for classical systems this has been done in \cite{Bi}).
 This would allow one to show  the relationship between long-time
 dynamics and free-energy landscape for quantum systems directly.
Finally, the precise definition of a ``quantum state''
is a delicate matter and merits further analysis. In this paper we have 
simply called ``state'' a minimum  of the TAP free-energy density.  One 
possible way to verify the existence and stability of these states 
is by studying the dynamics of this system starting from 
particular initial conditions as done in \cite{Alain,PaFr} for 
the classical model. This study is underway~\cite{Cugrsa1}.

\vskip 2cm

\noindent
{\bf Acknowledgments}
We wish to thank  D. R. Grempel, J. Kurchan,  
G. Lozano and C. A. da Silva Santos for very useful discussions. 
G. B. is supported by the Center of Material Theory, Rutgers University, 
NJ, USA. L. F. C. thanks ECOS-Sud for a travel grant and 
financial support from the grant ``Algorithmes d'optimisation et
syst{\`e}mes quantiques d{\'e}sordonn{\'e}s'', ACI-Jeunes Chercheurs, 2000.

\vskip 2cm

\noindent{\bf Appendix A}
\vspace{1cm}

In this appendix we show that the equation for $q_{\sc ea}$ that 
leads to a solution with the correct physical properties is 
Eq.~(\ref{eqq1final}) with the minus sign. 
Indeed, we search a solution that corresponds to a minimum of the 
TAP free-energy. The full stability analysis, that involves the 
evaluation of the complete Hessian of the TAP free-energy,  
is rather hard and has to be done numerically
(for instance, the form of $\tilde C(\omega)$
can only be obtained numerically).\\ 
However, we can still perform a partial analysis that suffices 
to justify the choice of the minus sign.  
Let us concentrate on the following diagonal elements of the Hessian:
\begin{eqnarray}
\label{q-stab}
\frac{\delta(-\beta F)}{\delta q_{\sc ea}^2}
&=&
-\frac{1}{2}\left[\left(1-\frac{p}{2}\right)\tilde{C}(0)+\frac{p}{2}\beta q_{\sc ea} 
\right]
\left[\frac{1}{(\tilde{C}(0)-\beta q_{\sc ea})^{2}q_{\sc ea}}-\frac{p(p-1)}{2}q_{\sc ea}^{p-3}
 \right]
\; ,
\\
\label{C(0)-stab}
\frac{\partial (-\beta F) }{\partial \tilde{C}(0) 
\partial \tilde{C}(0)}
&=&
\frac{-1}{(\tilde{C}(0)-q\beta )^{2}}
+\frac{p(p-1)}{2\beta }\int_{0}^{\beta } d\tau C^{p-2}(\tau)
\; .
\end{eqnarray}
From $z_\pm $'s definition we find 
\begin{equation}\label{ineg}
z_{-}\le \frac{-{\cal E}_{\sc th}}{p-1}\; , \qquad\qquad  
z_{+}\ge \frac{-{\cal E}_{\sc th}}{p-1} \; .
\end{equation}
Since $q_{\sc ea}$ is fixed by Eq.~(\ref{eqq1final}), 
the second factor on the 
right-hand-side of Eq.~(\ref{q-stab}) is positive (negative)
 for $z_{-}$ ($z_{+}$). A stable solution corresponds to a negative value
 of (\ref{q-stab}) and (\ref{C(0)-stab}),
 therefore one has to take the solution $q''$ for 
 $z_{-}$ and $q'$ for $z_{+}$. Moreover, 
since for ${\cal E}<{\cal E}_{\sc th}$ the right hand side of (\ref{eqq1final})
is positive then we obtain that $\tilde{C}(0)-\beta q'$ and 
$\tilde{C}(0)-\beta q''$ are positive and using that 
\[
\left(\frac{1}{\beta }\int_{0}^{\beta }C^{\alpha }(t)dt \right)\ge
\left(\frac{1}{\beta }\int_{0}^{\beta }C(t)dt \right)^{\alpha }
\qquad \alpha >1
\]
and imposing that (\ref{C(0)-stab}) has to be negative, 
we obtain:
\[
\frac{1}{(\tilde{C}(0)-\beta q_{\sc ea})
^{2}}-\frac{p(p-1)}{2}q_{\sc ea}^{p-2}\ge 0\qquad \mbox{for}\quad  q=q',q'' 
\] 
But this is impossible for $z_{+}$, i.e. it
 is not possible to have a consistent stable $z^{+}$ solution.

\vspace{1cm}
\noindent{\bf Appendix B}
\vspace{1cm}

In this Appendix we give an argument in favor of the  relation 
$\beta x=\partial
\sigma(\beta,f)/\partial f$  in Eq.~(\ref{mE}).
Let us start by assuming that it does hold and see that it 
leads to the equation linking $x$, $q_{\sc ea}$ and $C(\tau)$ in the 
Matsubara approach.
First of all we write the derivative of $\sigma $ with respect to 
$f$ as a derivative with respect to $\cal E$. This can be easily done
by  noticing that differentiating $f$ in Eq.~(\ref{eq:Gamma}) with respect to $\cal E$
 at $\beta$ and $\Gamma $ fixed is equivalent to differentiating $f$ 
with respect to $\cal E$ at $\beta ,\Gamma,q_{EA}$ and $C(\tau ) $ fixed
because $f$ is stationary in $q_{EA}$ and $C(\tau ) $. As a consequence we find
\begin{equation}
\beta x = \frac{\partial \sigma}{\partial f} =
\frac{\partial \sigma}{\partial {\cal E}} 
\frac{\partial {\cal E}}{\partial f} =
\frac{\partial \sigma}{\partial {\cal E}} \; q_{\sc ea}^{-p/2}
\; .
\end{equation}
The derivative $\partial \sigma/\partial {\cal E}$ can be 
easily computed from Eq.~(\ref{complexity}). One arrives at
\begin{equation}
\frac{p}{2} = \left( \tilde C(0)^2 -\beta^2 q_{\sc ea} (x-1) + \beta
q_{\sc ea} \tilde C(0) (x-2) \right)^{-1}
\; .
\end{equation}
In summary, starting from the TAP approach at fixed ${\cal E}$ we
obtain $q_{\sc ea}$ and $C(\tau)$. Assuming then that Eq.~(\ref{mE})
holds we obtain the equation linking $q_{\sc ea}$ and $x$ in the 
Matsubara approach. The equations for $C(\tau)$ in the TAP and 
Matsubara approaches, once $q_{\sc ea}$ is fixed, are identical. 
Hence we have proven that Eq.~(\ref{mE}) leads to the Matsubara 
results in \cite{Cugrsa}.  

The proof will be complete if we showed the other sense of the
implication.

\end{document}